\documentclass[graybox]{svmult}

\usepackage{mathptmx}       
\usepackage{helvet}         
\usepackage{courier}        
\usepackage{type1cm}        
\usepackage{makeidx}         
\usepackage{graphicx}        
\usepackage{multicol}        
\usepackage[bottom]{footmisc}

\makeindex             

\begin{document}

\title*{The significance of measurement independence for Bell inequalities and locality}

\titlerunning{Measurement independence, Bell inequalities, and locality}

\author{Michael J. W. Hall}
\institute{Centre for Quantum Computation and Communication Technology (Australian
Research Council), Centre for Quantum Dynamics, Griffith University, Brisbane, QLD 4111, Australia}
%
%
\maketitle

\abstract{A local and deterministic model of quantum correlations is always possible, as shown explicitly by Brans in 1988:  one simply needs the physical systems being measured to have a suitable statistical correlation with the physical systems performing the measurement, via some common cause.   Hence, to derive no-go results such as Bell inequalities, an assumption of measurement independence is crucial.  It is a surprisingly strong assumption -- less than $1/15$ bits of prior correlation suffice for a local model of the singlet state of two qubits -- with ramifications for the security of quantum communication protocols.  Indeed, without this assumption, any statistical correlations whatsoever -- even those which appear to allow explicit superluminal signalling -- have a corresponding local deterministic model.  It is argued that `quantum nonlocality' is bad terminology, and that measurement independence does not equate to `experimental free will'. Brans' 1988 model is extended to show that no more than $2\log d$ bits of prior correlation are required for a local deterministic model of the correlations between any two $d$-dimensional quantum systems.}

\section{Introduction}
\label{1}

Various ‘no-go’ results exist for models of quantum phenomena, based on
various more or less plausible assumptions for the structure of such models. Such
results support a longstanding view that quantum mechanics is more or less implausible  --- indeed, Niels Bohr was famously quoted as saying \cite{bohr}:
\begin{quotation}
Those who are not shocked when they first come across quantum theory cannot possibly have understood
it.
\end{quotation}
This has led not only to much philosophical discussion on which assumption(s) should be relaxed, but also to surprising applications of what might be termed ‘quantum implausibility’, such as quantum cryptography and
quantum computation.

The most remarkable of these no-go results are Bell inequalities, which imply that at least one of the plausible properties of determinism, locality and measurement independence must be given up to successfully describe quantum correlations between distant measurement regions \cite{bell,chsh,bell74}.  Here  measurement independence denotes the statistical independence of (i) any physical parameter influencing the selection of measurement procedures from (ii) any physical parameter influencing measurement outcomes, and is typically justified by an appeal to experimental free will  \cite{dialectica}.

The question of which property should be relaxed is not just a matter of idle speculation: the security of quantum cryptographic protocols, for example, relies on there being no deterministic description underlying correlations between distant measurement outcomes --- an eavesdropper possessing such a description would be able to determine the cryptographic key \cite{bellreview}.  Hence, any unconditionally secure protocol based on violation of Bell inequalities must, to ensure there is no deterministic description available, assume that the properties of locality and measurement independence hold. A similar requirement applies to device-independent protocols for randomness generation \cite{bellreview}.

While most discussion in the literature focuses on choosing between locality or determinism to model quantum correlations, it was pointed out by Brans in 1988 that there is in fact an explicit local {\it and} deterministic model, obtained by relaxing measurement independence \cite{brans}.  Brans further observed that there is an inherent conflict in assuming that both determinism and measurement independence hold: if the physical world is deterministic, then correlations between physical parameters are generic. 

There has been a recent upsurge of interest in local deterministic models, including their construction \cite{degorre, hallmodel,bg}; the derivation of generalised Bell inequalities incorporating a given level of measurement dependence  \cite{hallmodel, bg, relaxed,banik, koh,scarani, kay, gisinbell, chaves}; impacts on device-independent quantum communication protocols \cite{koh,scarani,kay}; and new experimental tests \cite{jason,gisinbellexp}.

In this contribution I briefly review the assumptions leading to  Bell inequalities (Sect.~\ref{2}); pause to urge replacement of the misleading terms `quantum nonlocality' and `Bell nonlocality' in the literature by the more neutral term `Bell nonseparability' (Sect.~\ref{3}); compare the degrees of measurement dependence of various local deterministic models for the singlet state, and extend the Brans model to show that local deterministic models for two $d$-dimensional quantum systems require no more than $2\log_2 d$ bits of measurement-dependent correlation (Sect.~\ref{4}); and discuss the relevance of local deterministic models to questions of causality and free will --- including a demonstration of the {\it prima facie }paradoxical existence of a local deterministic model for superluminal correlations (Sect.~\ref{5}).

\section{The well trod path to Bell inequalities}
\label{2}

\subsection{Bayes theorem}

Consider an experiment in which
\begin{itemize}
	\item Preparation procedure $P$ is carried out.
	\item Measurement procedure $M$ is performed.
	\item Outcome $m$ is recorded.
\end{itemize}
 In a joint measurement scenario one is interested in the case where the measurement procedure $M$ decomposes into two subprocedures $x$ and $y$, with respective outcomes $a$ and  $b$, i.e., 
 \[ M\equiv (x,y),\qquad m\equiv (a,b) . \] 
Statistical correlations between these outcomes are represented by some joint probability density $p(m|M,P)=p(a,b|x,y,P)$, which can be in principle measured via many repetitions of the experiment. Part of the physicist's job is to find an underlying model for these correlations, for a given set of experiments $\{(x,y,P)\}$.  

In particular, an underlying model introduces additional physical variables of some sort.  Denoting these underlying variables by $\lambda$,  Bayes theorem immediately tells us that
\begin{equation} \label{bayes}
p(m|M,P)=p(a,b|x,y,P) = \int \D\lambda \,p(a,b|\lambda,x,y,P) \, p(\lambda|x,y,P) ,
\end{equation}
with integration replaced by summation over any discrete ranges of $\lambda$.  A given model must therefore specify the type of information encoded in $\lambda$, and the underlying probability densities $p(a,b|\lambda,x,y,P)$ and $p(\lambda|x,y,P)$.

For example, in standard quantum mechanics $\lambda$ may be taken to range over a set of density operators, with
\begin{equation} \label{qm}
 p_Q(a,b|\lambda,x,y,P)={\rm tr}[\lambda E^{xy}_{ab}] ,\qquad p_Q(\lambda|x,y,P)=\delta(\lambda-\lambda_P),
\end{equation}
for some density operator $\lambda_P$ associated with preparation procedure $P$ and some positive operator valued measure (POVM) $\{E^M_m\equiv E^{xy}_{ab}\}$ associated with the joint measurement procedure $M=(x,y)$.

\subsection{Bell separability}
\label{2.2}

A given underlying model may or may not satisfy certain physically plausible properties, such as determinism, causal correlations, etc.  In the scenario typically considered for Bell inequalities \cite{bell}, one requires the quantities in Eq.~(\ref{bayes}) to satisfy the following three properties in particular:

\runinhead{Statistical completeness:} {\it All statistical correlations arise from ignorance of the underlying variable}, i.e., 
\begin{equation} \label{comp}
p(a,b|\lambda,x,y,P) = p(a|\lambda,x,y,P)\,p(b|\lambda,x,y,P) .
\end{equation}
Thus, all correlations between measurement outcomes vanish when the additional information encoded in $\lambda$ is specified \cite{jarrett}.  This property is also known as outcome independence, and is guaranteed to hold for {\it deterministic} models, i.e., for
\[p(a,b|\lambda,x,y,P,)\in\{0,1\}. \]
 Indeed, the existence of a deterministic model for a given set of correlations is equivalent to the existence of a statistically complete model \cite{relaxed,fine}. The original motivation for statistical completeness  was in fact outcome determinism, via an appeal to the existence of an underlying reality in which all measurement outcomes are predetermined~\cite{bell}.

\runinhead{Statistical locality:} {\it Distant measurement subprocedures do not influence each other's underlying outcome probability distributions}, i.e.,
\begin{equation} \label{local}
p(a|\lambda,x,y,P) = p(a|\lambda,x,P),\qquad p(b|\lambda,x,y,P) = p(b|\lambda,y,P) .
\end{equation}
Thus, an observer cannot distinguish, via any local measurement $x$, whether a distant observer has carried out measurement $y$ or $y'$, even given knowledge of the underlying variable $\lambda$.  This property, also known as parameter independence, is justified by the principle of relativity when the measurement subprocedures are carried out in spacelike separated regions \cite{bell}.  

\runinhead{Measurement independence:} {\it The measurement procedure $M=(x,y)$ is not correlated with the underlying variable}, i.e., 
\begin{equation} \label{mi}
p(\lambda|x,y,P) = p(\lambda|P).
\end{equation}
Thus, knowledge of the underlying variable gives no information about the measurement procedure, and vice versa.  This property is often justified by an appeal to `experimental free will' \cite{dialectica}, as will be discussed in some detail further below. 

The combination of all three properties is equivalent, via Eqs.~(\ref{bayes}) and (\ref{comp})-(\ref{mi}), to:
\begin{svgraybox} {\bf Bell separability:} The joint probabilities of distantly-performed measurement procedures have an underlying model of the form
\begin{equation} \label{bellsep}
p(a,b|x,y,P) = \int \D \lambda\, p(\lambda|P)\,p(a|\lambda,x,P) \,p(b|\lambda,y,P) ,
\end{equation}
i.e, a model satisfying statistical completeness, statistical locality and measurement independence.
\end{svgraybox}
Bell separable models were first introduced by John Bell \cite{bell} (see Ref.~\cite{wiseman} for a recent review), and capture the notion that statistical correlations between distant regions separate into independent contributions, distinguished by their dependencies on the measurement subprocedures as per Eq.~(\ref{bellsep}). 

\subsection{Bell inequalities}

Statistical correlations which have a Bell separable model satisfy various inequalities, known as Bell inequalities.  For example, if each measurement outcome is labelled by $\pm1$,  Bell separability implies that the Clauser-Horne-Shimony-Holt (CHSH) inequality \cite{chsh}
\begin{equation}
E(x,y,P) + E(x,y',P) + E(x',y,P) - E(x',y',P) \leq 2
\end{equation}
holds for any four pairs of distantly performed measurement procedures $(x,y)$, $(x,y')$, $(x',y)$ and $(x',y')$, where $E(x,y,P):=\sum_{a,b=\pm1} a\,b\,p(a,b|x,y,P)$.

It is now well known that not only do the predictions of standard quantum mechanics violate such Bell inequalities: so does nature \cite{aspect,hansen}.  Hence, our world is Bell-nonseparable, and one or more of the  three properties in Eqs.~(\ref{comp})-(\ref{mi}) must be relaxed in any underlying model thereof.

It is worth noting that the existence of a Bell separable model as per Eq.~(\ref{bellsep}), for some given set of joint measurement procedures $\{(x,y)\}$ and preparation procedure $P$, is also equivalent to the existence of a formal joint probability distribution for any finite subset $(x_1,\dots,x_m,y_1,\dots y_n)$ --- whether or not this subset has an experimental joint implementation.  In particular, one  may define \cite{fine}
\begin{eqnarray} \nonumber
p_F(a_1,\dots,a_m,b_1,\dots b_n|x_1,\dots,x_m,y_1,\dots y_n,P) := \int &\D \lambda& p(\lambda|P)\, \prod_{j=1}^m p(a_j|\lambda,x_j,P) \\
&~&~\times\prod_{k=1}^n p(b_k|\lambda,y_k,P) .
\end{eqnarray}
Bell inequalities correspond to boundary inequalities for the space (polytope) of such formal joint probability distributions \cite{bellreview,fineprl,garg,pit}.

\section{Why `quantum nonlocality' is bad terminology}
\label{3}

The violation of Bell inequalities by quantum systems is often referred to as `quantum nonlocality' or `Bell nonlocality'. To do so is quite misleading, however, as it implicitly --- and incorrectly --- suggests that some sort of mysterious action-at-a-distance is necessarily involved in quantum correlations.

In particular, only {\it one} of the three assumptions in Sect.~\ref{2.2} makes reference to notions of locality: {\it statistical locality} requires that a distant measurement procedure cannot be identified from local statistics. The other two assumptions, statistical completeness and measurement dependence, do not require any notion of acausal information transfer between distant regions (as shown explicitly in Sects.~\ref{4} and~\ref{5}).   Hence, by dropping either one of these two assumptions, Bell inequality violations may be modelled in a perfectly local manner.

Further, the property of statistical locality is automatically satisfied in standard quantum mechanics, via the usual tensor product representation
\begin{equation} \label{fact}
 E^{xy}_{ab} = E^x_a\otimes E^y_b 
 \end{equation}
of the POVM in Eq.~(\ref{qm}) for measurements in separated regions \cite{hall91}.  
Hence, quantum mechanics is in fact {\it local}, with respect to the only sense in which this concept makes an appearance in the derivation of Bell inequalities!  

This has led Mermin to conclude
that use of the term `quantum nonlocality' is no more
than ``fashion at a distance'' \cite{mermin}, and Kent to lament
it as a ``confusingly oxymoronic phrase'' that conflicts with
the notion of locality in quantum field theory \cite{kent}. An extended critique has  been given recently by \.Zukowksi and Brukner \cite{zb} (see also Ref.~\cite{hallgisin}).

While `Bell nonlocality' is preferable to `quantum nonlocality', insofar as the adjective vaguely implies some sort of special qualification, essentially the same criticisms apply. Moreover, since quantum communication protocols that rely on the violation of Bell inequalities {\it require} the assumption of statistical {\it locality} (and measurement independence), to ensure indeterminism (see Sect.~\ref{1}), it is similarly `confusingly oxymoronic' to assert that such protocols rely on Bell {\it nonlocality}, as is commonly done~\cite{bellreview}. 

I therefore strongly urge adoption of the more neutral term `Bell nonseparability'. 

\section{Local deterministic models of quantum correlations}
\label{4}

\subsection{Relaxing measurement independence}

As noted previously, violation of a Bell inequality, and hence of Bell separability, implies that at least one of the properties in Eqs.~(\ref{comp})-(\ref{mi}) must be relaxed in any underlying model. The degree to which these properties need to be individually or jointly relaxed, relative to various measures, has been recently reviewed \cite{relaxed}.

For example, the standard quantum mechanics model satisfies the property of statistical locality in Eq.~(\ref{local}), as noted in the previous section. Comparison of Eqs.~(\ref{qm}) and (\ref{mi}) shows that it also satisfies the property of measurement independence.  Hence, since it predicts violations of Bell inequalities, it follows that the standard quantum mechanics model must relax the property of statistical completeness. This is indeed so: the joint probability $p_Q(a,b|\lambda,x,y,P)$ in Eq.~(\ref{qm}) is only guaranteed to factorise, as per Eq.~(\ref{comp}), for tensor product states $\lambda=\lambda_1\otimes \lambda_2$.  

A major contribution by Brans was to provide, in contrast, the first explicit {\it local deterministic} model of quantum correlations, by instead relaxing the assumption of measurement independence \cite{brans}. The existence of such a fully causal model for Bell-nonseparable correlations further emphasises the point made in Sect.~\ref{3}, that the properties of statistical completeness and measurement independence do not rely on any concept of locality {\it per se}. 

Brans' model for the singlet state of two spin-$1/2$ 
particles or qubits, together with two subsequent models, are briefly described in Sect.~\ref{4.2}, before discussing the generalisation of Brans' model to arbitrary quantum correlations in Sect.~\ref{4.3}.  
However, it is of interest to first quantify the degree of measurement dependence of any given model, so as to be able to make quantitative comparisons between different models.

Several measures of measurement independence have been discussed in the literature \cite{hallmodel,bg,relaxed,banik,koh,scarani,kay,gisinbell}.  Attention here will be confined to the `measurement dependence capacity',
which directly quantifies the correlation between the joint measurement procedure and  underlying variable in terms of the maximum mutual information between them \cite{bg}:
\begin{equation} \label{cmd} 
C_{\rm MD} := \sup_{p(x,y)} H(\Lambda:X,Y) = \sup_{p(x,y)}\,\int\D\lambda\D x\D y\,p(\lambda,x,y|P)\log_2\frac{p(\lambda,x,y|P)}{p(\lambda|P)\,p(x,y)} .
\end{equation}
Here the supremum is over all possible probability densities $p(x,y)$ for the joint measurement procedure;  $H(\Lambda:X,Y)$ denotes the mutual information; and $p(\lambda,x,y,|P):=p(\lambda|x,y,P)\,p(x,y)$.  Note that the mutual information quantifies the average information gained about the measurement procedure from knowledge of the underlying variable, and vice versa, in terms of the number of bits required to represent the information \cite{inf}.  

The above measure vanishes if and only the measurement independence condition in Eq.~(\ref{mi}) is satisfied.  A useful upper bound follows via \cite{relaxed}
\begin{equation} \label{cbound}
C_{\rm MD} =\sup_{p(x,y)}\left[ H[\Lambda]-\int \D x\D y\,p(x,y) H_{x,y}(\Lambda)\right]\leq H_{\max} (\Lambda) - \inf_{x,y}\,H_{x,y}(\Lambda),
\end{equation}
where $H(\Lambda)$ denotes the  entropy of the underlying variable $\lambda$, with maximum possible value $H_{\max}(\Lambda)$, and $H_{x,y}(\Lambda)$ denotes the entropy of $p(\lambda|x,y,P)$.	
	
\subsection{Singlet state models}
\label{4.2}

\runinhead{Brans model:} Letting $P_S$ denote a preparation procedure for the singlet state $|\Psi_S\rangle$, the Brans model in its simplest form corresponds to choosing $\lambda\equiv (\lambda_1,\lambda_2)$, with $\lambda_1,\lambda_2=\pm1$, and identifying the labels $x$ and $y$ with measurement of spin in the corresponding unit directions $x$ and $y$.  The corresponding probabilities in Eq.~(\ref{bayes}) are then specified, via Eq.~(11) of Ref.~\cite{brans}, by
\begin{equation} \label{brans}
p_B(a,b|\lambda,x,y,P_S) :=\delta_{a,\lambda_1} \, \delta_{b,\lambda_2},\qquad p_B(\lambda|x,y,P_S):=\frac{1-\lambda_1\lambda_2 \,\,x \cdot y}{4} .
\end{equation}
This trivially reproduces the correct singlet state probabilities $p(a,b|x,y,P_S) = (1-ab\,x\cdot y)/4$, via Eq.~(\ref{bayes}). The model is deterministic, and clearly satisfies the properties of statistical completeness and statistical locality in Eqs.~(\ref{comp}) and (\ref{local}) --- but not the measurement independence property in Eq.~(\ref{mi}). Brans further showed that the correlation existing between the underlying variable $\lambda$ and the measurement procedures $x$ and $y$ can be simulated causally and deterministically \cite{brans}, as will be discussed in Sect.~\ref{5}.


To evaluate $C_{\rm MD}$ for this model, note first that $\lambda$ takes only 4 distinct values, implying that
 $H_{\max}(\Lambda)=\log_2 4 =2$ bits.  A straightforward calculation further gives
\[ H_{x,y}(\Lambda) = \log_2 2 +h(x\cdot y) \geq 1~{\rm bit}, \]
where $h(a)$ denotes the entropy of the probability distribution $\{(1\pm a)/2\}$.
Hence, using Eq.~(\ref{cbound}), the degree of measurement dependence in Eq.~(\ref{cmd}) is bounded by 1~bit of correlation.  It is straightforward to check that this bound is achieved by the choice $p(x,y)=[\delta(x+y)+\delta(x-y)]/(8\pi)$ in Eq.~(\ref{cmd}), where $x$ and $y$ range over all directions on the unit sphere, yielding
\begin{equation} \label{cdb}
C^{\rm B}_{\rm MD} = 1~{\rm bit}.
\end{equation}
Thus, no more than one bit of correlation is required in the Brans model for the singlet state.  This result is generalised in Sect.~\ref{4.3}.

\runinhead{Degorre {\it \bf et al.} model:}

The local deterministic model of the singlet state due to Degorre {\it et al.} takes the underlying variable $\lambda$ to be a point on the unit sphere, with~\cite{degorre}
\begin{equation} \label{degorre}
p_D(a,b|\lambda,x,y,P_S) :=\delta_{a,A(\lambda,x)} \, \delta_{b,B(\lambda,y)},\qquad p_D(\lambda|x,y,P_S) := \frac{|\lambda \cdot x|}{2\pi},
\end{equation}
where $A(\lambda,x):={\rm sign}\,\lambda\cdot x$ and $B(\lambda,y):=-{\rm sign}\,\lambda\cdot y$ determine the local outcomes. Thus, these outcomes correspond to the projections of `classical' spin vectors, $\lambda$ and $-\lambda$, onto the measurement directions $x$ and $y$ respectively. As in the Brans model, the properties of statistical completeness and statistical locality are clearly satisfied, while the property of measurement independence is clearly not. Note from Eq.~(\ref{degorre}) that $\lambda$ is only correlated with  {\it one} of the local measurement directions, $x$.  This model was also independently put forward by Barrett and Gisin \cite{bg}.

To calculate $C_{\rm MD}$ for the Degorre {\it et al.} model, note that the entropy of the underlying variable is maximised by a uniform distribution over the unit sphere, with $H_{\rm max}(\Lambda)=\log_2 4\pi$. Further, letting $\theta$ denote the angle between $\lambda$ and $x$, one has
\[
H_{x,y}(\Lambda) = -\int_0^\pi \D\theta\, \int_0^{2\pi}\D \phi \,\sin \theta\, \frac{|\cos\theta|}{2\pi} \log_2 \frac{|\cos\theta|}{2\pi} 
= \log_2 2\pi e^{1/2} .
\]
Hence, the upper bound in Eq.~(\ref{cbound}) is $\log_2 2/\sqrt{e}$.  This bound is achieved, for example, by $p(x,y)=(4\pi)^{-1}p(y)$, yielding \cite{bg}
\begin{equation} \label{cdeg}
C^{\rm D}_{MD} = \log_2 \frac{2}{\sqrt{e}} \approx 0.279~{\rm bits}.
\end{equation}
Thus, noting Eq.~(\ref{cdb}), the Degorre {\it et al.} model requires less correlation between the measurement directions and the underlying variable, in comparison to the Brans model.  Moreover, this correlation is only required between $\lambda$ and {\it one} of the local measurement directions.

\runinhead{Hall model:}

Finally, it is of interest to consider a local deterministic model with an even smaller degree of measurement dependence \cite{hallmodel}.  The underlying variable is again a point on the unit sphere, corresponding to a `classical' spin vector, and again the local outcomes are determined by the projection of this spin vector onto the local measurement directions, via
\begin{equation}
p_H(a,b|\lambda,x,y,P_S) :=\delta_{a,A(\lambda,x)} \, \delta_{b,B(\lambda,y)} ,
\end{equation}
with  $A(\lambda,x)={\rm sign}\,\lambda\cdot x$ and $B(\lambda,y)=-{\rm sign}\,\lambda\cdot y$  as for the Degorre {\it et al.} model.
However, in contrast to the latter model,
\begin{equation}
p_H(\lambda|x,y,P_S) := \frac{1}{4\pi} ~ \frac{1+(x\cdot y) \,{\rm sign}\,\left[(\lambda\cdot x)(\lambda\cdot y)\right] }{1 +(1-2\phi_{xy}/\pi)\,{\rm sign}\,\left[(\lambda\cdot x)(\lambda\cdot y)\right] },
\end{equation}
where $\phi_{xy}\in[0,\pi]$ denotes the angle between directions $x$ and $y$.

The interesting aspect of this model is its low degree of measurement dependence, with \cite{relaxed}
\begin{equation}
C^{\rm H}_{MD} \approx 0.0663~{\rm bits}.
\end{equation}
Thus, in comparison to Eqs.~(\ref{cdb}) and (\ref{cdeg}), a remarkably low degree of correlation, less than $1/15$ of a bit, is required to model the singlet state.

\subsection{Generalising the Brans model to arbitrary quantum states}
\label{4.3}

While the Brans model of the singlet state is not optimal with respect to the degree of measurement dependence required, it does have the significant advantage of being easily generalisable to a local deterministic model for {\it all} quantum correlations, with a corresponding simple upper bound for the degree of measurement dependence required.


In particular, consider a preparation procedure $P$ corresponding to some quantum density operator $\rho$ describing two quantum systems, where these systems have a $d_1$-dimensional and a $d_2$-dimensional Hilbert space respectively. Further, consider an arbitrary joint measurement of two Hermitian observables, $x$ and $y$, on these systems, with corresponding POVM $\{ E^{xy}_{ab}\}$. The joint outcome $(a,b)$ may be labelled by elements of the set $O:=\{1,2,\dots,d_1\}\times\{1,2,\dots,d_2\}$, without any loss of generality.
To construct a corresponding local deterministic model, we choose the underlying variable to be  $\lambda=(\lambda_1,\lambda_2) \in O$, and generalise Eq.~(\ref{brans}) to
\begin{equation}
p_B(a,b|\lambda,x,y,P):=\delta_{a,\lambda_1} \, \delta_{b,\lambda_2},\qquad p_B(\lambda|x,y,P):=  {\rm tr}[\rho E^{xy}_{\lambda_1\lambda_2}]
\end{equation}
Substitution into Eq.~(\ref{bayes}) immediately recovers the quantum probability density $p(a,b|x,y,P)= {\rm tr}[\rho E^{xy}_{ab}]$, as required.

To obtain an upper bound for the degree of measurement dependence required in this model, note that $0\leq H(\Lambda)\leq \log_2 d_1d_2$, with the upper bound corresponding to a uniform distribution over $\lambda$.  Hence, from Eq.~(\ref{cbound}), one immediately has the upper bound
\begin{equation}
C^{\rm B}_{\rm MD} \leq  \log_2 d_1 + \log_2 d_2
\end{equation}
for the degree of measurement independence.  In particular, no more than $2\log_2 d$ bits of correlation are required to model the statistics of all Hermitian observables on two $d$-dimensional quantum systems.  Note that  the bound holds irrespective of whether the corresponding POVMs factorise as per Eq.~(\ref{fact}). It would be of interest to determine whether the bound also holds for non-Hermitian observables, i.e., for arbitrary joint POVMs.


\section{Questions of causality and free will}
\label{5}

\subsection{Causality in measurement dependent models}
\label{5.1}

Brans did more than give the first explicit local and deterministic model for singlet state correlations.  He also showed that the corresponding probability distribution $p_B(\lambda|x,y,P)$ was compatible with a fully causal explanation, and that it did not contradict the notion of `experimental free will' in any operational sense \cite{brans}.  The causal aspect will be discussed here, and the free will aspect in Sect.~\ref{5.2}. The surprising existence of a local deterministic model for {\it superluminal} correlations is given in Sect.~\ref{5.3}.

The violation of measurement independence, i.e., a correlation such that
\[ p(\lambda|x,y,P)\neq p(\lambda|P), \]
may at first sight suggest that the joint measurement procedure $(x,y)$ has a causal effect on the statistics of $\lambda$.  However, this is not so: using Bayes rule the above equation can equivalently be written in either of the forms
\[ p(x,y|\lambda,P)\neq p(x,y|P),\qquad p(\lambda,x,y|P)\neq p(\lambda|P)\,p(x,y|P),  \]
where the latter form is seen to be perfectly symmetrical with respect to the measurement procedure and $\lambda$. Correlation does not specify causation.

In fact, for any measurement-dependent correlation, $p_0(\lambda|x,y,P)\neq p_0(\lambda|P)$ say, and any distribution of joint measurement procedures, $p_0(x,y|P)$ say, the corresponding probability density $p_0(x,y|\lambda,P)\neq p_0(x,y|P)$ is uniquely determined by the laws of probability. It is straightforward to construct a causal model for this probability density, of the form 
\begin{equation} \label{causal}
p_0(x,y|\lambda,P) = \int \D\mu \,p(x|\mu)\,p(y|\mu)\,p(\mu|\lambda,P) ,
\end{equation}
where $\mu$ is a further underlying variable.  It is clear from this equation that the correlation can be causally implemented via generation of the distribution of $\mu$ by $\lambda$ and $P$, with subsequent local generation of the distribution of the measurement subprocedures $x$ and $y$ by $\mu$, with no retrocausal or superluminal propagation required. 

As an explicit example, choose
$\mu\equiv (\mu_1,\mu_2)$, where $(\mu_1,\mu_2)$ labels the set of possible joint measurement procedures $\{(x,y)\}$, with $p(x|\mu):= \delta(x-\mu_1)$,  $p(y|\mu):=\delta(y-\mu_2)$, and
\begin{equation}   
p(\mu|\lambda,P):= \frac{p_0(\lambda|\mu_1,\mu_2,P)\,p_0(\mu_1,\mu_2|P)}{\int\!\int \D \mu_1 \D \mu_2 \, p_0(\lambda|\mu_1,\mu_2,P)\,p_0(\mu_1,\mu_2|P)} .
\end{equation}
It is straighforward to check that these choices reproduce Eq.~(\ref{causal}), as desired. Thus no violation of causality is required by measurement dependent models, such as those in Sect.~\ref{4}.


\subsection{Free will and conspiracy}
\label{5.2}

Brans noted that the assumption of measurement independence is fundamentally inconsistent with a fully deterministic world \cite{brans}.  In such a world even the preparation procedure $P$, along with the measurement $M=(x,y)$ and the outcomes $m=(a,b)$, will be predetermined by suitable underlying variables, and hence are  generically all correlated.  Thus, in a superdeterministic world, assuming that $x$ and $y$ are {\it not} correlated with the underlying variables, i.e., measurement independence, amounts to conspiracy!  In such a world, measurement dependence --- and hence the possibility of Bell inequality violation --- is only to be expected~\cite{thooft}.

However, as previously remarked, measurement independence is typically justified by an appeal to `experimental free will': surely the selection of measurement procedures is  independent of any underlying physical variables that determine the outcomes?  For if they were not, surely this would compromise our perception of having the free will to make such a selection?

There are several responses that can be made in this regard, in addition to the obvious point that the subjective experience of free will does not imply its objectivity.  The first is practical: in actual tests of Bell nonseparability and Bell inequalities, physical systems rather than physicists are used to `randomly' select measurement procedures \cite{aspect, hansen}. This fact has practical relevance for the security of commercial quantum cryptographic devices that contain such systems: how can we trust random number generators built by a third party? \cite{koh,scarani,kay}.  We certainly do not expect these devices to have `free will', and their degree of measurement dependence is easily manipulated by the device manufacturer.

The second is operational: there is no experimental distinction that can be made between models satisfying measurement independence and models which do not \cite{brans} --- at least, under the proviso that the distribution of measurement procedures is independent of the preparation procedure, $p(x,y|P)=p(x,y)$.  In such a case, all that is operationally accessible to the `free will' of the experimenter(s) in this regard is the choice of $p(x,y)$.  However, {\it any} such choice is compatible with measurement-dependent models: it merely implies that the operationally-inaccessible joint distribution $p(\lambda, x,y|P)$ is given by $p(\lambda|x,y,P)\,p(x,y)$ \cite{relaxed}.

The third is rhetorical: suppose that experimenters were informed that there was a physical quantity they could not change: no matter what choices of preparation and measurement procedures they made, using their `free will', the quantity mysteriously came out to be the same --- even for joint measurements in spacelike separated regions.  Would this necessarily represent a lack of `free will'? No, not if the quantity was the total energy! Conservation laws are not considered to be conspiratorial.  This suggests the intriguing possibility  of a local deterministic model for quantum systems in which $p(x,y)$ emerges as a conserved quantity \cite{dilorenzo}.

Thus, there is no {\it a priori} reason why the behaviour of experimenters or random generators should {\it not} be statistically correlated with a given system to some degree, reflecting a common causal dependence on some underlying variable, even in the absence of superdeterminism and/or in the presence of `free will'.  However, it must be admitted that a measurement-dependent model in which $p(x,y)$ emerged as a conserved quantity would be far more compelling than those presented in Sect.~{\ref{4}.

\subsection{Local deterministic models of \underline{ superluminal} correlations}
\label{5.3}

It is remarkable to note that even correlations which appear to allow superluminal signalling can be modelled in a local and deterministic manner, by relaxing the assumption of measurement independence.

For example, consider some preparation procedure $P$ and two joint measurement procedures $M=(x,y)$ and $M'=(x',y)$, with corresponding experimental joint probability distributions satisfying
\begin{equation} \label{sig} 
p(b|x,y,P) \neq p(b|x',y,P) .
\end{equation}
Thus, knowledge of the local outcome distribution of procedure $y$ provides information about whether the procedure $x$ or $x'$ was performed.  This is an example of  a `signalling' correlation. Such a correlation is not surprising  in the case that the measurement subprocedure $y$ is performed in the future lightcone of subprocedure $x$ or $x'$ --- this would simply represent the possiblity of signalling from the past to the future.  However, such a correlation would be very surprising in the case of spacelike-separated subprocedures $x$ and $y$, as it would appear to amount to the possibility of superluminal signalling.

Surprisingly, perhaps, a local deterministic model for such signalling correlations is easily obtained, via a straightforward adaptation of the extended Brans model discussed in Sect.~\ref{4.3}. In particular, for a given set of experimental joint probability distributions $\{p_E(a,b|x,y,P)\}$, choose the underlying variable to range over the set of possible joint measurement outcomes, with  $\lambda=(\lambda_1,\lambda_2)\in \{(a,b)\}$, and define
\begin{equation}
p(a,b|\lambda,x,y,P) :=\delta_{a,\lambda_1} \, \delta_{b,\lambda_2},\qquad p(\lambda|x,y,P):= p_E(\lambda_1,\lambda_2|x,y,P).
\end{equation}
This model is explicitly deterministic, and clearly satisfies both statistical completeness and statistical locality, {\it whether or not} the experimental correlations are signalling!  Further, a causal description of the measurement-dependent correlation $p(\lambda|x,y,P)$ can always be given, as per the discussion in Sect.~\ref{5.1} above.

The resolution of this paradoxical result is that the very notion of `signalling' logically requires some degree of measurement independence: if one has, for example, no control at all over the choice of measurement $x$ or $x'$ in Eq.~(\ref{sig}), then one has no ability at all to signal --- e.g., `buy' or `sell' --- via such a choice \cite{scarani}.  

It follows that one should not refer to `no-signalling' or `signal locality' without a simultaneous commitment to measurement independence (I must recant having done so previously \cite{relaxed}).  Moreover, a simple tweak of the above model implies that it is possible to replace  any underlying model that violates the property of statistical locality by one that instead violates measurement independence.  Indeed, in this regard Barrett and Gisin have previously shown that any underlying deterministic model that requires at most $m$ bits of superluminal communication can be replaced by one that requires $C_{MD}\leq m$ bits of measurement-dependent correlation \cite{bg}.

\section{Conclusions}

One of the most remarkable discoveries in physics is the violation of Bell separability by quantum phenomena: any underlying model of such phenomena must relax at least one of the properties of statistical completeness, statistical locality or measurement independence.  There is a strong intuition among physicists that perfect correlations between distant measurement outcomes, such as singlet state correlations, should be deterministically and locally mediated, independently of the joint measurement procedure. However, this intuition fails in the light of Bell inequality violation.  

Given that standard quantum mechanics satisfies statistical locality and measurement independence, Occam's razor suggests that it is the intuition behind determinism (and thus statistical completeness) that must be given up.  On the other hand, it may be argued that relaxing  measurement dependence is relatively far more efficient: only $1/15$ of a bit of measurement dependence is required to model the singlet state, in comparison to 1 bit of communication in nonlocal models, and 1 bit of shared randomness in nondeterministic models \cite{relaxed}. In the end, however, whether or not one's personal preference is guided by simplicity or efficiency, the consideration of all three properties cannot be avoided --- and is of practical relevance in assessing the reliability of device-independent quantum communication protocols. 

It is a pleasure to be able to acknowledge the seminal contribution of Carl Brans to this ongoing debate, as part of this Festschrift to mark his 80th birthday.  His explicit local deterministic model for quantum correlations has led to a better understanding of the significance of measurement (in)dependence, and has stimulated many new results and ideas. He will no doubt be pleased that one of the latter \cite{jason} may lead to an experimental connection with his many cosmological interests: the recent proposal to test measurement independence in a Bell inequality experiment by using the light from distant quasars that have never been in causal contact.

\begin{acknowledgement}
	I thank  Cyril Branciard, Antonio Di Lorenzo, Jason Gallicchio, Nicolas Gisin, Valerio Scarani, Joan Vaccaro, Mark Wilde and Howard Wiseman for various stimulating discussions on this topic over the past few years. This work was supported by the ARC Centre of Excellence CE110001027.
\end{acknowledgement}

\end{document}